\documentclass[journal]{IEEEtran}
\usepackage{amsmath,amsfonts}
\usepackage{amsthm}
\usepackage{algorithm}
\usepackage{algpseudocode}
\usepackage{array}
\usepackage[caption=false,font=normalsize,labelfont=sf,textfont=sf]{subfig}
\usepackage{textcomp}
\usepackage{stfloats}
\usepackage{url}
\usepackage{verbatim}
\usepackage{graphicx}
\usepackage{cite}%\usepackage[backend=bibtex]{biblatex}
\usepackage{bm}
\usepackage{color}
\usepackage{cuted}
\usepackage{algpseudocode}
\newcounter{MYtempeqncnt}

\begin{document}

\title{\huge Sub-6GHz Assisted mmWave Hybrid Beamforming with Heterogeneous Graph Neural Network}

\author{Zhaohui Huang, Zhaocheng Wang, \emph{Fellow, IEEE}, and Sheng Chen, \emph{Life Fellow, IEEE} %
\thanks{This work was supported in part by the National Natural Science Foundation of China under Grant U22B2057, in part by Guangdong Optical Wireless Communication Engineering and Technology Center, in part by Shenzhen VLC System Key Laboratory, and in part by Shenzhen Solving Challenging Technical Problems (JSGG20191129143216465) \emph{(Corresponding author: Zhaocheng Wang)}.} %
\thanks{Z. Huang, and Z. Wang are with Department of Electronic Engineering, Tsinghua University, Beijing 100084, China. Z. Wang is also with the Shenzhen International Graduate School, Tsinghua University, Shenzhen 518055, China (E-mails: hzh21@mails.tsinghua.edu.cn, zcwang@tsinghua.edu.cn).} %
\thanks{Sheng Chen is with School of Electronics and Computer Science, University of Southampton, Southampton SO17 1BJ, U.K. (E-mail: sqc@ecs.soton.ac.uk).} %
\vspace*{-2mm}
}

%\markboth{XXXX}%
%{\MakeLowercase{\textit{et al.}}:}

\maketitle

\begin{abstract}
In next-generation communications, sub-6GHz and millimeter-wave (mmWave) links typically coexist, with the sub-6GHz link always active and the mmWave link active when high-rate transmission is required. Due to the spatial similarities between sub-6GHz and mmWave channels, sub-6GHz channel information can be utilized to support hybrid beamforming in mmWave communications to reduce overhead costs. We consider a multi-cell heterogeneous communication network where both sub-6GHz and mmWave communications co-exist. Multiple mmWave base stations (BSs) in the heterogeneous network simultaneously transmit signals to multiple users in their own mmWave cells while interfering with each other. The challenging problem is to design hybrid beamformers in the mmWave band that can maximize the system spectral efficiency. To address this highly complex programming using sub-6GHz information, a novel heterogeneous graph neural network (HGNN) architecture is proposed to learn the intrinsic relationship between sub-6GHz and mmWave and design the hybrid beamformers for mmWave BSs. The proposed HGNN consists of two different node types, namely, BS nodes and user equipment (UE) nodes, and two different edge types, namely, desired link edge and interfering link edge. In addition, the attention mechanism and the residual structure are utilized in the HGNN architecture to improve the performance. Simulation results show that the proposed HGNN can successfully achieve better performances with sub-6GHz information than traditional learning methods. The results also demonstrate that the attention mechanism and residual structure improve the performances of the HGNN compared to its unmodified counterparts.
\end{abstract}


\section{Introduction}\label{Introduction}

Because of its huge accessible bandwidth, millimeter-wave (mmWave) transmission has been identified as a major approach for next-generation wireless communications \cite{Bai}. To overcome the high path loss of mmWave signals, mmWave communication systems typically employ huge antenna arrays and directional beamforming/precoding \cite{Zpi}. However, huge antennas result in high power consumption of radio frequency (RF) components in fully digital baseband precoding modules and high overhead costs caused by channel state information (CSI) estimation for beamforming/precoding.

In order to reduce the power consumption of RF components while achieving satisfactory performance, the hybrid beamforming technique is utilized in the mmWave MIMO systems \cite{Sohrabi,Li}. Typically, either fully-connected or partially-connected architectures are adopted, depending on whether each RF chain is connected to all antennas (the former) or to a disjoint subset of antennas (the latter) \cite{Ayach,Sorhrabi}. There are several techniques to obtain the hybrid beamforming configuration \cite{Shi,Sun,Alkhateeb2,Zhu,Chen}. The traditional methods, such as \cite{Shi,Alkhateeb2,Zhu}, usually rely on optimization theory to derive optimal or near optimal iterative algorithms for solutions, which impose high computational complexity and require the whole antenna array's CSI or the optimal fully digital beamforming configuration. To reduce the computational complexity, the work \cite{Chen} proposed a deep unfolding based architecture, which replaces some original components with their learnable counterparts, and achieves faster convergence than traditional methods. However, these techniques all require the entire antenna array's CSI, leading to high overhead cost.

To reduce the overhead of establishing a mmWave link, various approaches exploiting channel sparsity were proposed \cite{Alkhateeb,Choi,Dong}. These methods focus on obtaining a compressed sensing channel estimation that trades the measurement overhead for performance. Another technique to reduce the training overhead is to leverage the out-of-band information extracted from low-frequency channels \cite{Ali,Gao}. Experiments in \cite{Ali,Peter} demonstrated the feasibility of using the low-frequency information due to the similarities of spatial characteristics of sub-6 GHz and mmWave channels. The study \cite{Ali} also showed that the power azimuth spectrums (PASs) are almost consistent in low-frequency and mmWave channels. In addition, the work \cite{Jiang} showed the close relationship of the power delay profiles (PDPs) between the sub-6GHz channel and the mmWave channel. With the aforementioned similar characteristics of low-frequency and mmWave channels, the authors of \cite{Nithsche} demonstrated the construction of the overhead-free multi-Gbps mmWave link with out-of-band inference, while the authors of \cite{Hashemi} showed the selection of the mmWave beam based on the estimated line-of-sight (LOS) direction of sub-6GHz channel. In addition, machine learning methods were also utilized to take advantage of the out-of-band information \cite{Ma,Gao,Alrabeiah}. For example, the work \cite{Ma} developed a dedicated deep learning model based on 3-dimensional (3D) convolutional neural network (CNN) to provide the mmWave beam selection with the aid of low-frequency information. The authors of \cite{Gao} proposed a dual-input neural network to predict the optimal beam using both the sub-6GHz channel and a few pilots in the mmWave band transmitted from a few active antennas. The work \cite{Alrabeiah} proved the existence of the mapping functions that can predict the optimal mmWave beam and blockage status directly from the sub-6GHz channel under certain conditions, and developed a deep learning model that can predict the optimal mmWave beam and blockage with high success probability. However, the application of \cite{Gao, Ma, Alrabeiah} is limited to the single-user scenario.

It is challenging to efficiently utilize the sub-6GHz CSI in mmWave hybrid beamforming, because to our knowledge the correlation between sub-6GHz and mmWave has not be derived in closed-form formulas, and hybrid beamforming itself is usually a non-convex problem due to mutual interference among links, coupling, and constraints of optimization variables.

Graph neural network (GNN) \cite{GNN} provides a structural learning framework for graph-based problems. Through convolution operations between adjacent nodes, GNN can extract local and global information on the graph data, achieving classification, regression or prediction tasks. From a theoretical perspective, there are three advantages of GNNs compared to CNNs and multilayer perceptions (MLPs) in many communication problems.

\begin{enumerate}
\item Scalability and parallelization: The wireless communication network can be naturally modeled as a graph, where nodes correspond to base stations (BSs) and user equipment (UEs) with edges representing the channels between them. The graph convolution structure of GNN ensures its adaptability to input graph data with varying numbers of nodes (i.e., UEs and/or BSs), whereas MLP and CNN are constrained by strict dimensional requirements for their inputs. In addition, GNNs allow for parallel computation on graph data, contributing to a more efficient runtime performance.
\item Permutation invariance (PI) and permutation equivalence (PE): A function or model has PI property if its output remains unchanged when the order of its input elements is altered. Conversely, a function or model has PE property if the order of its outputs is correspondingly altered when its input elements are permuted. The PI and PE properties are proved to be universal in many wireless communication problems, such as power allocation, beamforming, and interference mitigation. The GNN can achieve PI or PE properties, distinguishing itself from CNN and MLP, which lack these properties. Consequently, the GNN emerges as a more suitable solution for numerous wireless communication problems.
\item Requiring less training samples: It is theoretically proved that GNNs require fewer training samples than MLPs to achieve the same performance, and the training-samples demand gap grows with the number of nodes in the graph \cite{Shen_Theory}. A reduced number of training samples means lower overhead and less computational complexity, and therefore the GNNs are advantageously in wireless communication problems.
\end{enumerate}

Empirically, compared with other deep learning neural networks (DNNs), such as CNN and MLP, GNN provides high performance in various wireless communication problems, such as power allocation, beamforming, reconfigurable intelligent surface (RIS) configurations \cite{Shen,Guo,Zhang,Chowdhury}. For example, the work \cite{Shen} proposed a wireless channel graph convolutional network (WCGCN) to solve the power allocation and beamforming problem in device-to-device (D2D) communication systems. The study \cite{Guo} proposed a parameter sharing structure as a heterogeneous GNN (HGNN) for learning power control in cellular systems. Moreover, the authors of \cite{Chowdhury} unfolded a power allocation enabled iterative weighted minimum mean squared error (WMMSE) algorithm with a distributed GNN architecture, which reduces the computational complexity and has robustness and generalizability in different densities and sizes. The research \cite{Chowdhury} also showed the relationship between the GNN and deep unfolding techniques.

For heterogeneous wireless networks, the GNNs with some specific modifications, such as HGNNs and bipartite GNNs (BGNNs), offer advantages over traditional methods or other deep learning methods. For example, the work \cite{XCZhang} proposed an unsupervised HGNN to solve the power control/beamforming in heterogeneous D2D networks, showing the applicability of dedicated GNNs in heterogeneous networks. In addition, the authors of \cite{Zhang} utilized the GNN to appropriately schedule users and design RIS configurations to achieve high overall throughput while considering fairness among the users, demonstrating the potential of GNNs in asymmetric communication problems. Furthermore, the study [30] proposed a BGNN to solve the scalable multi-antenna beamforming optimization problems, exhibiting the GNN's flexibility with respect to the system size, i.e., the number of antennas or users.

Inspired by these related works on GNNs for heterogeneous wireless networks, in this paper, we propose a HGNN, which consists of two different node types, namely, BS nodes and UE nodes, and two different edge types, namely, desired link edge and interfering link edge. Our HGNN achieves the fully-connected or partially-connected hybrid beamforming configurations in mmWave with the aid of both sub-6GHz CSI and partial mmWave CSI.
We utilize this HGNN to solve the challenging hybrid beamforming problem in heterogeneous mmWave and sub-6GHz networks. The main contributions are summarized as follows.

\begin{enumerate}
\item The hybrid beamforming problem is formulated to maximize the spectral efficiency of the mmWave system in the heterogeneous cellular network (HCN) with power constraints. We model the HCN as a heterogeneous graph, where BSs and UEs are modeled as nodes, while desired links and interfering links represent the edges between the corresponding nodes. Based on this heterogeneous graph, the HGNN is proposed to solve the fully-connected or partially-connected hybrid beamforming problem. To reduce the overhead, we leverage the sub-6GHz CSI and the partial mmWave CSI in solving this optimization design.

\item To improve the performance of the proposed HGNN, the attention mechanism and the residual structure are utilized in the GNN structure. The attention mechanism is added in the aggregation procedure to adaptively learn the importance of different messages. The residual structure is added in the combination procedure to eliminate the degradation phenomenon. The introduction of attention mechanism and residual structure does not affect the PI and PE properties as well as scalability of GNNs. 

\item Numerical results verify that the proposed HGNN outperforms other machine learning methods in various scenarios. Moreover, the utilization of attention mechanism and residual structure is shown to enhance the achievable performance. Besides, the strong scalability and low running complexity of the proposed HGNN are demonstrated.
\end{enumerate}

The rest of this paper is organized as follows. Section~\ref{system_model} describes the heterogeneous mmWave and sub-6GHz network and formulates the system spectral efficiency maximization problem. The proposed HGNN is detailed in Section~\ref{architecture}. In Section~\ref{sim}, numerical results are presented to verify the effectiveness of the proposed HGNN. Finally, Section~\ref{clu} concludes the paper.

\emph{Notations}: Scalars, vectors and matrices are represented by normal face lowercase letters, boldface lowercase letters and boldface uppercase letters, respectively, e.g., $a$, $\mathbf{a}$ and $\mathbf{A}$. $\mathbb{C}^{m\times n}$ and $\mathbb{R}^{m\times n}$ denote the $m$ by $n$ dimensional complex space and real space, respectively. $\mathbf{I}_k$ is the $k\times k$ identity matrix, and $\textsf{j} = \sqrt{-1}$ is the imaginary axis. The transpose and Hermitian transpose are denoted by $(\cdot)^{\rm T}$ and $(\cdot)^{\rm H}$, respectively. The complex normal distribution is represented by $\mathcal{CN}(\mu,\sigma^2)$ with mean $\mu$ and variance $\sigma^2$. $\mathbb{E}(\cdot)$ and $\|\cdot\|_F$ are the expectation and Frobenius norm, respectively, while $|\mathcal{A}|$ is the cardinality of the set $\mathcal{A}$. The $m$th-row and $n$th-column entry of $\mathbf{A}$ is denoted by $\mathbf{A}_{[m,n]}$. The concatenation of two vectors $\mathbf{a}$ and $ \mathbf{b}$ is given by $[\mathbf{a}\, \Vert\, \mathbf{b}]=\big[\mathbf{a}^{\rm T}, \mathbf{b}^{\rm T}\big]^{\rm T}$.

\begin{figure}[t!]
\vspace*{-2mm}
\begin{center}
		\includegraphics[width=1.0\columnwidth]{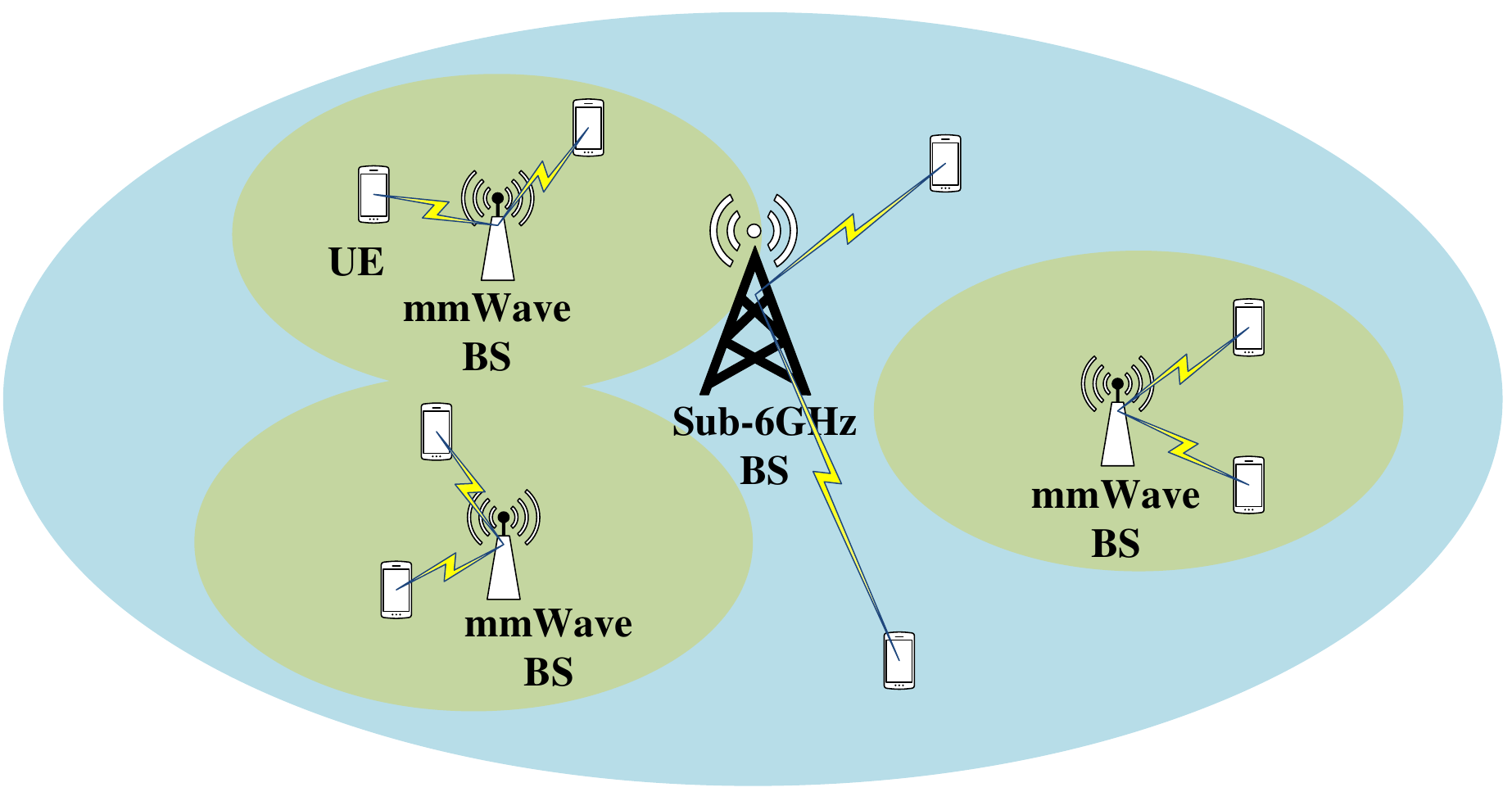}
\end{center}
\vspace*{-2mm}
\caption{The heterogeneous mmWave and sub-6GHz network.}
\label{mmWaveNetwork} % Fig.1
\vspace*{-2mm}
\end{figure}

\section{System model and problem formulation}\label{system_model}

\begin{figure}[t]
\begin{center}
\includegraphics[width=0.9\columnwidth]{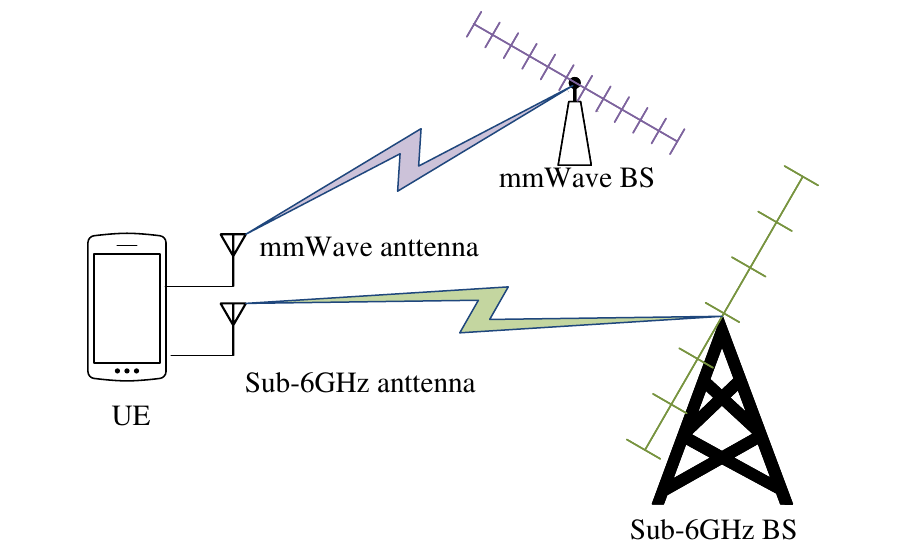}
\end{center}
\vspace{-2mm}
\caption{UE communicates over both sub-6GHz and mmWave bands with corresponding BSs.}
\label{UE} % Fig.2
\vspace*{-2mm}
\end{figure}

Consider a heterogeneous sub-6GHz and mmWave communication system, with $K$ BSs equipped with $N_m$ mmWave antennas and one BS equipped with $N_s$ sub-6GHz antennas, as illustrated in Fig.~\ref{mmWaveNetwork}. The coverage of sub-6GHz signals is the entire network while the coverage of mmWave signals is only in the corresponding mmWave cell. The sub-6GHz antenna array is fully digital, with each antenna connected to an independent RF chain, and the mmWave antenna array is a hybrid architecture. All the users are equipped with both a sub-6GHz antenna and a mmWave antenna, as shown in Fig.~\ref{UE}. Both sub-6GHz and mmWave communications are considered under single-carrier system for simplicity. Nevertheless, our proposed method can be easily extended to an OFDM system. It is assumed that all the UEs are constantly connected to the central sub-6GHz BS in sub-6GHz, and the mmWave connection is only active when a high transmission rate is required. For simplicity, we omit the sub-6GHz links in Fig.~\ref{mmWaveNetwork} for the UEs with both mmWave and sub-6GHz links. MmWave BS $k$, $k=1,2,\cdots,K$, can serve up to $I_k$ users in cell $k$ with mmWave channel. The total number of UEs is hence given by $I_{sum}=\sum_{k=1}^K I_k$. The $i$th user in mmWave cell $k$ is denoted as $i_k$. Our goal is to find the mmWave hybrid beamforming configurations by utilizing sub-6GHz CSI and some estimation of partial mmWave CSI. The sub-6GHz beamforming can be handled by the WMMSE \cite{Shi} or other traditional methods.

\subsection{Hybrid beamforming in the transmitter at mmWave band}\label{S2.1}

We assume that mmWave BS $k$ with $N_{RF,k}$ RF chains communicates with each UE via only one stream. Thus, the number of data streams of BS $k$ is $I_k$. Moreover, the number of RF chains of BS $k$ is usually larger than the number of users that can be served simultaneously by BS $k$, i.e., $I_k\leq N_{RF,k}$. For simplicity, we also assume that BS $k$ will utilize the $I_k$ RF chains to serve the corresponding $I_k$ UEs.

Let $\mathbf{s}_{k}=\left[s_{k}[1],s_{k}[2],\cdots,s_{k}[I_k]\right]^{\rm T}\in\mathbb{C}^{I_k\times 1}$ denote the transmitted symbols of BS $k$, where $\mathbb{E}\big(\mathbf{s}_k\mathbf{s}_k^{\rm H}\big) = \mathbf{I}_{I_k}$ and $s_{k}[i]$ is the transmitted data for UE $i_k$. Assuming that the hybrid precoder of BS $k$ is $\mathbf{F}_k\in\mathbb{C}^{N_m \times I_k}$, the precoded signal of BS $k$ is given by
\begin{equation}\label{eq1}
	\mathbf{x}_k = \mathbf{F}_k \mathbf{s}_k.
\end{equation}
The hybrid precoder $\mathbf{F}_k = \mathbf{F}_{RF,k} \mathbf{F}_{BB,k}$ is composed of the analog precoder $\mathbf{F}_{RF,k}\in\mathbb{C}^{N_m\times I_k}$ and the baseband precoder $\mathbf{F}_{BB,k} \in \mathbb{C}^{I_k \times I_k}$. We denote $\mathbf{F}_{k} = \left[\mathbf{f}_{k}[1],\mathbf{f}_{k}[2],\cdots,\mathbf{f}_{k}[I_k]\right]$ and $\mathbf{F}_{RF,k} = \left[\mathbf{f}_{RF,k}[1],\mathbf{f}_{RF,k}[2],\cdots,\mathbf{f}_{RF,k}[I_k]\right]$.

\begin{figure}[!t]
\vspace*{-2mm}
\begin{center}
\includegraphics[width=1.0\columnwidth]{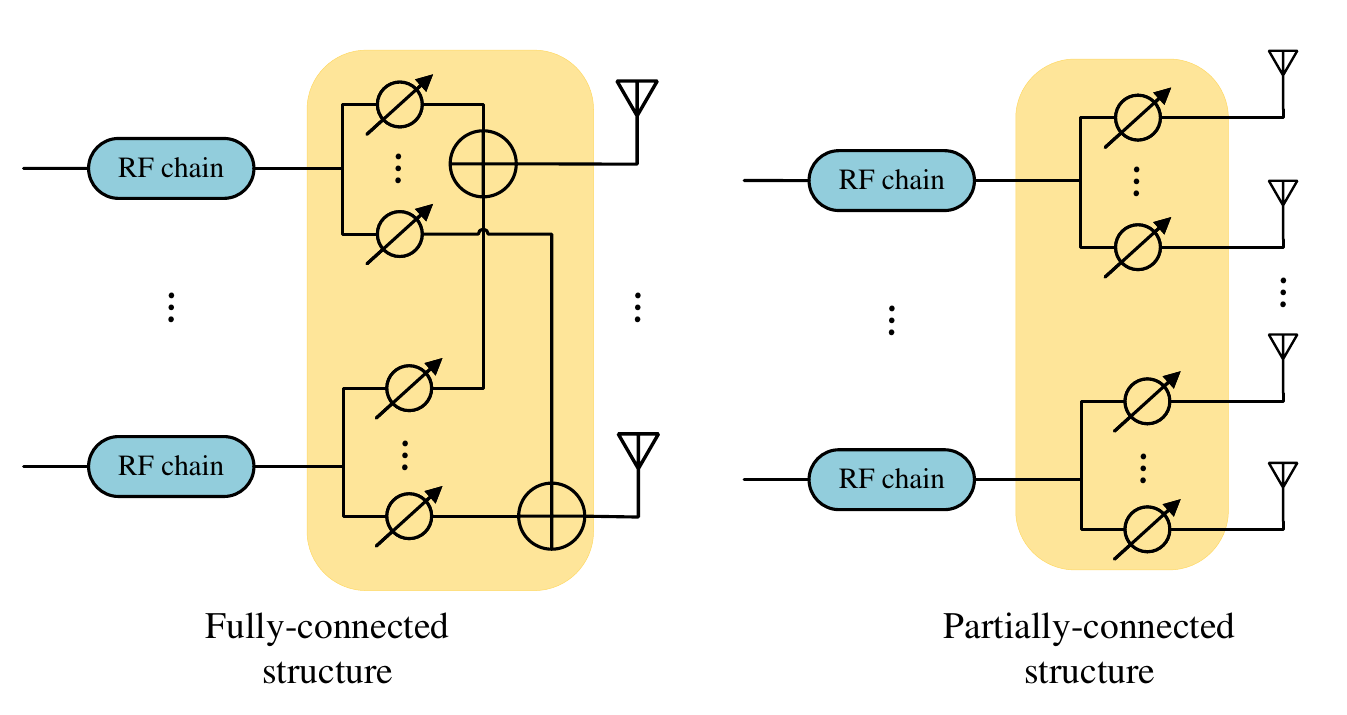}
\end{center}
\vspace{-2mm}
\caption{The fully-connected and partially-connected hybrid beamforming structures in the transmitter.}
\label{hybrid} % Fig.3
\vspace{-2mm}
\end{figure}

In our system model, we design GNNs for both fully-connected hybrid beamforming structure and partially-connected structure, which are shown in Fig.~\ref{hybrid}.

\subsubsection{Fully-connected structure}

In the fully-connected hybrid beamforming structure \cite{Alkhateeb3}, each RF chain is connected to all antennas via phase shifters, which implies that each entry of $\mathbf{F}_{RF,k}$ has a constant modulus. All the entries are normalized to satisfy $\left|{\mathbf{F}_{RF,k}}_{[m,n]}\right|^2 = \frac{1}{N_m}$, i.e., ${\mathbf{F}_{RF,k}}_{[m,n]}=\frac{1}{\sqrt{N_m}}e^{\textsf{j}\varphi_{k,m,n}}$, where $\varphi_{k,m,n}\in \mathbb{C}$ represents the phase of the $m$-th phase shifter in the $n$-th RF chain of the $k$-th BS. $\mathbf{F}_{BB,k}$ is forced to satisfy $\big\|\mathbf{F}_{RF,k}\mathbf{F}_{BB,k}\big\|^2_F \leq P_k$, where $P_k$ is the maximum transmitting power of BS $k$.

\subsubsection{Partially-connected structure}

The partially-connected structure, also known as array of subarray structure, uses significantly fewer phase shifters for energy efficiency \cite{LDai,SHan,XYu}. This structure connects each RF chain output signal to $N_m/N_{RF,k}$ antennas, which reduces the RF hardware complexity. The analog precoder $\mathbf{F}_{RF,k}$ can be written as a block diagonal matrix as 
\begin{equation}\label{Fpartially} % eq.2
	\mathbf{F}_{RF,k} = 
    \begin{pmatrix}
        \underline{\mathbf{f}}_{1,k} & \underline{\mathbf{0}} & \cdots & \underline{\mathbf{0}} \\
        \underline{\mathbf{0}} & \underline{\mathbf{f}}_{2,k} & \cdots & \underline{\mathbf{0}} \\
        \vdots & \vdots & \ddots & \vdots\\
        \underline{\mathbf{0}} & \underline{\mathbf{0}} & \cdots & \underline{\mathbf{f}}_{N_{RF,k},k} \\
    \end{pmatrix},
\end{equation}
where the row vector $\underline{\mathbf{f}}_{i,k}\! =\! \frac{1}{\sqrt{N_m}}\left[e^{\textsf{j}\varphi_{k,(i-1)\frac{N_m}{N_{RF,k}}+1}}, \cdots, e^{\textsf{j}\varphi_{k,i\frac{N_m}{N_{RF,k}}}} \right]\! \in\! \mathbb{C}^{1\times \frac{N_m}{N_{RF,k}}}$ represents the phases of the $i$-th RF chain at the $k$-th BS, and $\underline{\mathbf{0}}\in \mathbb{C}^{1\times \frac{N_m}{N_{RF,k}}}$ is the zero row vector. Similar to the fully-connected case, $\mathbf{F}_{BB,k}$ is forced to satisfy the power constraint of $\big\|\mathbf{F}_{RF,k}\mathbf{F}_{BB,k}\big\|^2_F \leq P_k$.

The received signal $y_{i_k}$ at the receiver $i_k$ can be formulated as 
\begin{align}\label{y_i_k} % eq.3
	y_{i_k} =& \sum_{m=1}^{K} \mathbf{h}_{i_k,m}^{\rm H} \mathbf{x}_m + n_{i_k} \nonumber \\ 
	=& \mathbf{h}_{i_k,k}^{\rm H} \mathbf{f}_k[i_k] \mathbf{s}_k[i_k] + \sum_{l=1,l\neq i_k}^{I_k} \mathbf{h}_{i_k,k}^{\rm H} \mathbf{f}_k[l] \mathbf{s}_k[l] + \nonumber \\
	 &\sum_{m=1,m\neq k}^{K}\sum_{l=1}^{I_m}\mathbf{h}_{i_k,m}^{\rm H} \mathbf{f}_m[l] \mathbf{s}_m[l] + n_{i_k},
\end{align}
where $\mathbf{h}_{i_k,k}\in \mathbb{C}^{N_m \times 1}$ denotes the channel from BS $k$ to UE $i_k$, $n_{i_k}$ denotes the additive white Gaussian noise (AWGN) with the distribution $\mathcal{CN}(0,\sigma_{i_k}^2)$. On the right hand side of the last equation of (\ref{y_i_k}), the four terms are the desired signal, intracellular interference, intercellular interference and noise, respectively.

\subsection{MmWave channel model}\label{S2.2}

The mmWave channel from BS $k$ to UE $i_k$ can be formulated as \cite{Gao}
\begin{equation}\label{eqCh} % eq.4
	\mathbf{h}_{i_k,k} = \sqrt{\frac{N_m}{N_c}}\sum_{l=1}^{N_{c}}\beta_{l}e^{\textsf{j}(\theta_l+2\pi \tau_l B)}\mathbf{a}(\phi_l),
\end{equation}
where $N_c$ is the number of paths, $B$ is the bandwidth, $\beta_l$, $\theta_l$ and $\tau_l$ are the attenuation coefficient, phase and propagation delay of the $l$-th path, respectively, while $\phi_l$ is the
angle of departure (AoD) of the $l$-th path, and $\mathbf{a}(\phi_l)$, $1\le l\le N_c$, are the steering vectors at the departure side. In (\ref{eqCh}), all the index marks of BS $k$ and UE $i_k$ are omitted for clarity. For simplicity, we assume that each BS is equipped with a uniform linear array (ULA) for mmWave communications. Therefore, the array response vector can be formulated as
\begin{equation}\label{eqAR} % eq.5
\mathbf{a}(\phi) = \frac{1}{\sqrt{N_m}}\left[1,e^{\textsf{j}\frac{2\pi}{\lambda}d \sin(\phi)},\cdots,e^{\textsf{j}\frac{2\pi}{\lambda}(N-1) d \sin(\phi)}\right]^{\rm T},
\end{equation}
where $\lambda$ is the signal wavelength and $d$ is the antenna spacing, which is set to $d=\lambda/2$.

\subsection{Problem formulation}\label{S2.3}

Our target is to design the hybrid precoders to maximize the spectral efficiency of the whole system. Let $R_{i_k}$ be the capacity of UE $i_k$, which is formulated as (\ref{R_i_k}) shown at the top of the next page.
Then, the spectral efficiency maximization problem can be formulated as (\ref{max}) shown at the top of the next page, where the constraint (\ref{7a}) is the transmitting power constraint for each BS and the constraint (\ref{7b1}) or (\ref{7b2}) is the constant modulus constraint for the phase shifters.

\begin{figure*}[!t]
	% ensure that we have normalsize text
	\normalsize
	% Store the current equation number.
	\setcounter{MYtempeqncnt}{\value{equation}}
	% Set the equation number to one less than the one
	% desired for the first equation here.
	% The value here will have to changed if equations
	% are added or removed prior to the place these
	% equations are referenced in the main text.
	\setcounter{equation}{5}
	\begin{equation}\label{R_i_k} % eq.6
R_{i_k} = \log_2\left(1 + \frac{\|\mathbf{h}_{i_k,k}^{\rm H} \mathbf{f}_k[i_k]\|^2}{\sum_{l=1,l\neq i_k}^{I_k}\|\mathbf{h}_{i_k,k}^{\rm H} \mathbf{f}_k[l]\|^2 + \sum_{m=1,m\neq k}^{K}\sum_{l=1}^{I_m}\|\mathbf{h}_{i_k,m}^{\rm H} \mathbf{f}_m[l]\|^2 +\sigma_{i_k}^2}\right)
	\end{equation}
	\begin{align}\label{max} % eqs.7, 7a, 7b1, 7b2
		\max\limits_{\forall \mathbf{F}_{RF,k},\mathbf{F}_{BB,k}} & \sum_{k=1}^{K}\sum_{i=1}^{I_k}R_{i_k}  \\
		\text{s.t.}\hspace*{5mm} & \big\|\mathbf{F}_{RF,k}\mathbf{F}_{BB,k}\big\|_F^2\leq P_k, \forall{k} \tag{7a}\label{7a} \\
		& \text{for fully-connected structure: } \big\|{\mathbf{F}_{RF,k}}_{[m,n]}\big\|^2=\frac{1}{N_m}, \forall{k,m,n} \tag{7b1} \label{7b1} \\
		& \text{for partially-connected structure: } \mathbf{F}_{RF,k} \text{ defined in } (\ref{Fpartially})  \tag{7b2} \label{7b2}
	\end{align}
	% Restore the current equation number.
	\setcounter{equation}{\value{MYtempeqncnt}}
	% The IEEE uses as a separator
	\hrulefill
	% The spacer can be tweaked to stop underfull vboxes.
	\vspace*{4pt}
\end{figure*}

Due to the power constraint (\ref{7a}) and the constant modulus constraint (\ref{7b1}) or (\ref{7b2}), the problem (\ref{max}) is nonconvex, which is challenging to solve. Moreover, we want to solve this nonconvex problem not relying on the full mmWave CSI but relying on the sub-6GHz CSI and some estimation of the partial mmWave CSI, which makes the problem even more challenging.

\subsection{Partial mmWave CSI}\label{S2.4}

For simplicity, we assume that the perfect sub-6GHz CSI is available. Denote the sub-6GHz channel from BS $0$ to UE $i_k$ as $\widetilde{\mathbf{h}}_{i_k} \in \mathbb{C}^{N_s \times 1}$. Some works \cite{Sim,Ren} showed the similarities between sub-6GHz channel and mmWave channel and the feasibility of predicting the optimal mmWave channel via sub-6GHz CSI in the single user scenario. But the sub-6GHz CSI alone, which does not contain direct interference information or desired link information between BSs without sub-6GHz antennas and their serving UEs, is completely insufficient to optimize the problem (\ref{max}). However, it is expensive to obtain the full mmWave CSI, since the full CSI estimation at mmWave band requires changing the phase shifters $\mathbf{F}_{RF,k}$ or baseband precoders $\mathbf{F}_{BB,k}$ at the transmitter $N_m$ times, which imposes huge overhead cost.

Therefore, we follow \cite{Gao} to use the partial channel information on $\bar{N}_m \le N_m$ antennas at mmWave band, namely, partial mmWave CSI, to assist the hybrid precoding. In the training process, $\bar{N}_m$ mmWave antennas are active while the others are inactive for each BS. In addition, all the RF chains are only connected to the active antennas through the corresponding phase shifters. Denote the partial channel from BS $k$ to UE $i_k$ as $\bar{\mathbf{h}}_{i_k,k}$. The partial mmWave CSI offers a rough description of the full mmWave CSI and can provide the candidate strong directions for the whole mmWave channel \cite{Ali}, which motivates us to adopt the partial mmWave CSI for hybrid precoding. The channel estimation of $\bar{\mathbf{h}}_{i_k,k}$ can be achieved via the least squares (LS) or linear minimum mean squared error (LMMSE) methods with much fewer pilots than what needed for the estimation of the full mmWave channel $\mathbf{h}_{i_k,k}$. We adopt the partial mmWave CSI and sub-6GHz CSI to solve the optimization problem (\ref{max}).

\section{Proposed HGNN}\label{architecture} % S3

\subsection{Heterogeneous graph representation of heterogeneous network}\label{hg} % S3.1

According to \cite{Sun2}, an information graph can be defined as a directed graph $G = (\mathcal{V},\mathcal{E})$ with a node type mapping function $\tau:\mathcal{V}\rightarrow\mathcal{A}$ and a link type mapping function $\phi:\mathcal{E}\rightarrow\mathcal{R}$, where each node $\mathbf{v}\in\mathcal{V}$ belongs to a particular node type $\tau(\mathbf{v})\in\mathcal{A}$, each link $\mathbf{e}\in\mathcal{E}$ belongs to a particular relation $\phi(\mathbf{e})\in\mathcal{R}$, and if two links belong to the same relation type, the two links share the same starting node type as well as the same ending node type. If the types of nodes $|\mathcal{A}|> 1$ or the types of relations $|\mathcal{R}|>1$, the network is called a heterogeneous network; otherwise it is a homogeneous network. The neighborhood of a node $\mathbf{v}_m$ is defined as $\mathcal{N}(\mathbf{v}_m)=\{\mathbf{v}_n|\mathbf{e}_{m,n}\in \mathcal{E}\}$. In addition, we define the subset of the neighborhood for node $\mathbf{v}_m$ with node type $\tau(\mathbf{v}_n)=t$ by $\mathcal{N}_t(\mathbf{v}_m)=\{\mathbf{v}_n|\mathbf{e}_{m,n}\in \mathcal{E}, \tau(\mathbf{v}_n)=t\}$, and the subset of the neighborhood $\mathcal{N}_{t,w}(\mathbf{v}_m)=\{\mathbf{v}_n|\mathbf{e}_{m,n}\in \mathcal{E},\phi(\mathbf{e}_{m,n})=w, \tau(\mathbf{v}_n)=t\}$ is defined as the subset of $\mathcal{N}_t(\mathbf{v}_m)=\{\mathbf{v}_n|\mathbf{e}_{m,n}\in \mathcal{E}, \tau(\mathbf{v}_n)=t\}$ where all links have the same relation $\phi(\mathbf{e}_{m,n})=w$.

\begin{figure}[t]
\vspace*{-2mm}
\begin{center}
\includegraphics[width =1.0\columnwidth]{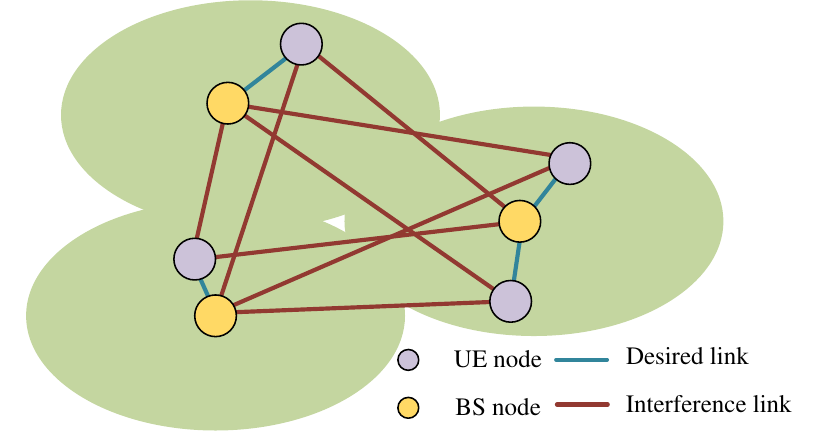}
\end{center}
\vspace{-2mm}
\caption{The heterogeneous graph representation of the communication system described in Section~\ref{system_model}.}
\label{HetGraph} % Fig.4
\vspace*{-2mm}
\end{figure}

As illustrated in Fig.~\ref{HetGraph}, it is natural to model the heterogeneous mmWave and sub-6GHz cellular network described in Section~\ref{system_model} as a heterogeneous graph. Specifically, there are two different node types, namely, BS node and UE node, and two different edge types, namely, desired link and interfering link. The desired links represent the communication links at mmWave band from BS $k$ to its serving UEs $i_k$, while the interfering links represent the interfering links at mmWave band from BS $k$ to its non-serving UEs $i_{k'}, k'\neq k$. The links at sub-6GHz band are omitted in this graph. Since each UE has a sub-6GHz link to the BS with sub-6GHz antennas, the sub-6GHz link features can be included in the UE node features. 

In this heterogeneous graph model for the system described in Section~\ref{system_model}, the node feature for BS $k$, which is denoted as $v_{k}=P_k\in \mathbb{R}^{1\times 1}$, consists of the maximum power of BS $k$, while the node feature vector of UE $i_k$, which is denoted as $\mathbf{v}_{i_k}=\big[\sigma^2_{i_k},\widetilde{\mathbf{h}}_{i_k}^{\rm T}\big]^{\rm T}\in \mathbb{C}^{(1+N_c)\times 1}$, consists of the noise power and the sub-6GHz CSI from the central (sub-6GHz) BS to UE $i_k$. The edge feature of the link from BS $k$ to UE $i_k$ consists the partial mmWave CSI, i.e., $\mathbf{e}_{i_k,k}=\bar{\mathbf{h}}_{i_k,k}\in\mathbb{C}^{\bar{N}_m\times 1}$. The node types are $\mathcal{A}=\{\text{b(BS)}, \text{u(UE)}\}$, and the edge types are $\mathcal{E}=\{\text{d(desired link)}, \text{i(interfering link)}\}$.

\subsection{Architecture of the proposed HGNN}\label{S3.2}

\begin{figure*}[t!]
\vspace*{-2mm}
\begin{center}
\includegraphics[width=2.0\columnwidth]{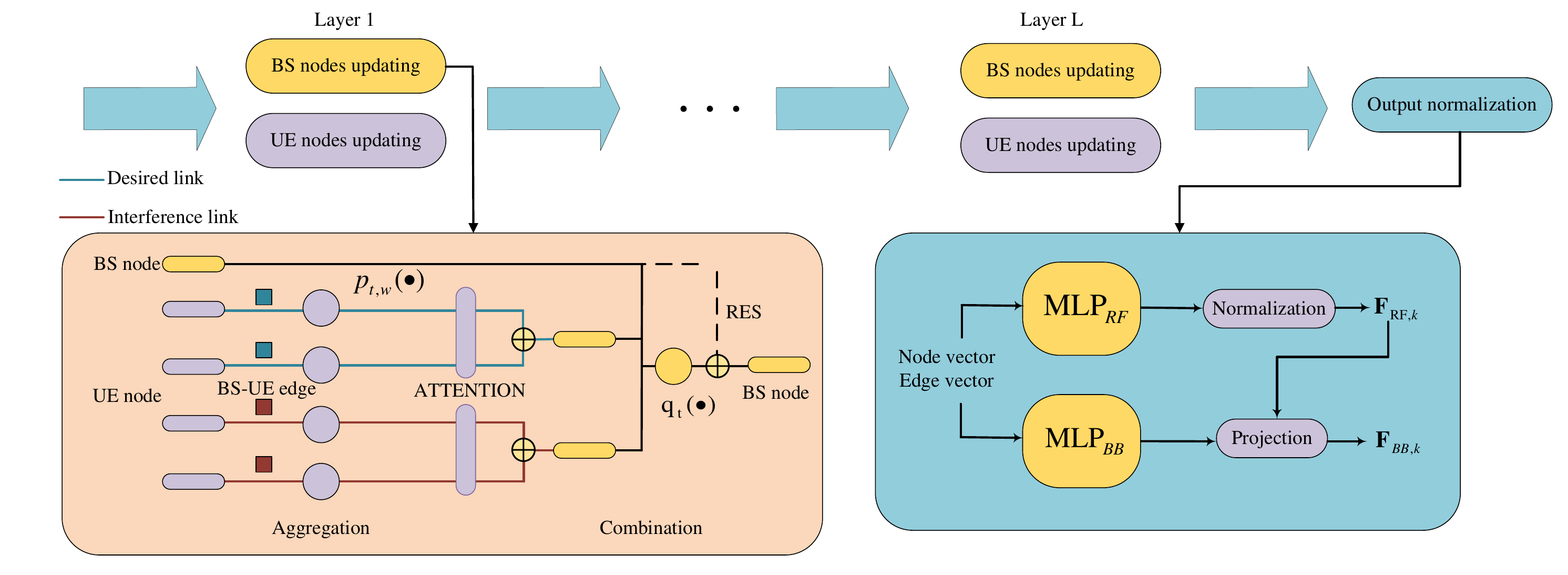}
\end{center}
\vspace*{-2mm}
\caption{The architecture of the proposed HGNN.}
\label{GNN} % Fig.5
\vspace*{-2mm}
\end{figure*}

In this subsection, we detail the proposed HGNN for solving the optimization problem (\ref{max}). The architecture of the HGNN, depicted in Fig.~\ref{GNN}, consists of three main components: 1)~attention based aggregation, 2)~res-based combination, and 3)~output normalization.

\subsubsection{Attention based aggregation}

The aggregation procedure is utilized to aggregate the information of neighboring nodes and edges. In the traditional HGNN, the aggregated outputs of the $t\in \mathcal{A}$ type of nodes and $w\in \mathcal{E}$ type of edges for node $\mathbf{v}_i$ in layer $l$ are formulated as \cite{Guo}
\setcounter{equation}{7}
\begin{equation}\label{agg} % eq.8
	\mathbf{a}_{i,t,w}^{(l)} = AGG_{j\in\mathcal{N}_{t,w}(\mathbf{v}_i)}\left(p_{t,w}\big(\mathbf{v}_j^{(l-1)},\mathbf{e}_{i,j}\big)\right),
\end{equation}
where $\mathbf{v}_j^{(l-1)}$ denotes the node features in layer $l-1$, $AGG_{j\in\mathcal{N}_{t,w}(\mathbf{v}_i)}(\cdot)$ denotes the aggregated function and $p_{t,w}(\cdot)$ is an MLP, which is identical for all node-edge pairs with node type $t$ and edge type $w$. $AGG_{j\in\mathcal{N}_{t,w}(\mathbf{v}_i)}(\cdot)$ is a function that satisfies the commutative law and is usually adopted as summation or maximization.

In the traditional aggregation procedure (\ref{agg}), the information of all the neighboring nodes and edges are aggregated with the same importance. However, in the heterogeneous graph, the importance of different messages may be different. Therefore, we further incorporate the attention mechanism into the aggregated function $AGG_{j\in\mathcal{N}_{t,w}\big(\mathbf{v}_i^{(l-1)}\big)}(\cdot)$. The attention function maps the inputs, a query and a set of key-value pairs, onto an output, which is computed as a weighted sum of the input values, where all the query, keys and values are vectors \cite{Vaswani}. In the aggregated function $AGG_{j\in\mathcal{N}_{t,w}(\mathbf{v}_i)}(\cdot)$, query is set as $\mathbf{v}_i^{(l-1)}$, keys and values are set as $p_{t,w}(\mathbf{v}_j^{(l-1)},\mathbf{e}_{i,j})$. By adopting attention, (\ref{agg}) can be rewritten as
\begin{equation}\label{aggWat} % eq.9
	\mathbf{a}_{i,t,w}^{(l)} = \sum_{j\in\mathcal{N}_{t,w}(\mathbf{v}_i)} \alpha_{i,j,t,w}\cdot p_{t,w}\big(\mathbf{v}_j^{(l-1)},\mathbf{e}_{i,j}\big),
\end{equation}
where the weights $\alpha_{i,j,t,w}$, which satisfy $0\leq\alpha_{i,j,t,w}\leq 1$ and $\sum_{j\in\mathcal{N}_{t,w}(\mathbf{v}_i)}\alpha_{i,j,t,w}=1$, represents the importance of the neighborhood $\mathbf{v}_j$ to the original node $\mathbf{v}_i$, and they are calculated as (\ref{eqATws}) shown at the top of the page, where $\mathrm{ReLU}(x)=\max\{x,0\}$ is the activation function and $\mathbf{att}_{t,w}$ are the trainable parameters in the attention function.

\begin{figure*}[!t]
	% ensure that we have normalsize text
	\normalsize
	% Store the current equation number.
	\setcounter{MYtempeqncnt}{\value{equation}}
	% Set the equation number to one less than the one
	% desired for the first equation here.
	% The value here will have to changed if equations
	% are added or removed prior to the place these
	% equations are referenced in the main text.
	\setcounter{equation}{9}
\begin{equation}\label{eqATws} % eq.10
	\alpha_{i,j,t,w} = \frac{\exp\left(\mathrm{ReLU}\left(\mathbf{att}_{t,w}^{\rm T} \big[\mathbf{v}_i^{(l-1)} \, \Vert \, \mathbf{v}_j^{(l-1)}\, \Vert \, \mathbf{e}_{i,j}\big]\right)\right)}{\sum_{k\in\mathcal{N}_{t,w}(\mathbf{v}_i)}\exp\left(\mathrm{ReLU}\left(\mathbf{att}_{t,w}^{\rm T}\big[\mathbf{v}_i^{(l-1)} \, \Vert \, \mathbf{v}_k^{(l-1)}\, \Vert \, \mathbf{e}_{i,k}\big]\right)\right)}
\end{equation}
\setcounter{equation}{12}
\begin{equation}\label{eqMLPrfF} % eq.13
	{\mathbf{F}_{RF,k}}_{[i,j]} = \frac{{MLP_{RF}\big(\mathbf{v}_k^{(L)}\big)}_{[i,j]}}{\sqrt{N_m}\left|{MLP_{RF}\big(\mathbf{v}_k^{(L)}\big)}_{[i,j]}\right|},  1\leq i\leq N_m, 1\leq j\leq N_{RF,k}
\end{equation}
\begin{equation}\label{eqMLPrfP} % eq.14
	{\mathbf{F}_{RF,k}}_{[i,j]} = \left\{ \begin{array}{cl}
		\frac{{MLP_{RF}\big(\mathbf{v}_k^{(L)}\big)}_{[i,j]}}{\sqrt{N_m}\left|{MLP_{RF}\big(\mathbf{v}_k^{(L)}\big)}_{[i,j]}\right|}, & (j-1)\frac{N_m}{N_{RF,k}} + 1 \leq i\leq j\frac{N_m}{N_{RF,k}}, 1\leq j\leq N_{RF,k} \\
		0, & \text{otherwise} \\
	\end{array}\right.
\end{equation}

	% Restore the current equation number.
	\setcounter{equation}{\value{MYtempeqncnt}}
	% The IEEE uses as a separator
	\hrulefill
	% The spacer can be tweaked to stop underfull vboxes.
	\vspace*{4pt}
\end{figure*}

\subsubsection{Res-based combination}

The combination procedure is utilized to combine the node information with the aggregated information from its neighborhood. In the traditional HGNN, the combination outputs in layer $l$ are formulated as \cite{Guo}
\setcounter{equation}{10}
\begin{equation}\label{comb} % eq.11
	\mathbf{v}_i^{(l)} = COMB_{\tau(\mathbf{v}_i)}\left(\mathbf{v}_i^{(l-1)},\big\{\mathbf{a}_{i,t,w}^{(l)},t\in\mathcal{A},w\in\mathcal{E}\big\}\right),
\end{equation}
where $COMB_{\tau(\mathbf{v}_i)}(\cdot)$ represents the combination function for node type $\tau(\mathbf{v}_i)$ with trainable parameters.

The traditional combination procedure (\ref{comb}) is a simple concatenation of the node features and the aggregated features. However, when the layers of the HGNN are deep, this structure of combination may suffer from the severe degradation problem \cite{He} and the over-smoothing problem where the outputs of all nodes may have the same features \cite{Oono}. Therefore, we adopt the residual structure \cite{He} to solve the degradation problem. The output of the res-based structure is a summation of the input and output of the original network. Therefore, with the residual structure, (\ref{comb}) is rewritten as 
\begin{equation}\label{combWres} % eq.12
	\mathbf{v}_i^{(l)} = \mathbf{v}_i^{(l-1)} + q_{\tau(\mathbf{v}_i)}\left(\mathbf{v}_i^{(l-1)},\big\{\mathbf{a}_{i,t,w}^{(l)},t\in\mathcal{A},w\in\mathcal{E}\big\}\right),
\end{equation}
where $q_{\tau(\mathbf{v}_i)}(\cdot)$ with $\tau(\mathbf{v}_i)=t$ is an MLP with trainable parameters.

\subsubsection{Output normalization}

After $L$ layers of aggregation and combination, the node features of the last layer are fed into an output normalization layer to obtain the solutions of the problem (\ref{max}), which consists of two MLPs, $MLP_{RF}(\cdot)$ and $MLP_{BB}(\cdot)$, for obtaining the analog beamforming matrix $\mathbf{F}_{RF,k}$ and the baseband precoding matrix $\mathbf{F}_{BB,k}$, respectively. 

\emph{3.1)~Analog beamforming solution:} For the fully-connected structure, the MLP $MLP_{RF}(\cdot)$ with a normalization module is utilized to obtain $\mathbf{F}_{RF,k}$ with input $\mathbf{v}_k^{(L)}$ according to (\ref{eqMLPrfF}) shown at the top of previous page. For the partially-connected structure, the normalization module is only forced on the block diagonal elements and the other elements are set to zero, which can be formulated as (\ref{eqMLPrfP}) shown at the top of previous page.

\emph{3.2)~Baseband precoding solution:}  After obtaining $\mathbf{F}_{RF,k}$, the baseband precoding matrix is generated as follows. First, the MLP $MLP_{BB}(\cdot)$ is utilized to produce:
\setcounter{equation}{14}
\begin{equation}\label{eqMLPbbI} % eq.15
	\mathbf{f}'_{BB,k}[i_k] = MLP_{BB}\left([\mathbf{v}_k^{(L)}\,\Vert\,\mathbf{v}_{i_k}^{(L)}]\right).
\end{equation}

Denote $\mathbf{F}'_{BB,k} = \left[\mathbf{f}'_{BB,k}[1],\mathbf{f}'_{BB,k}[2],\cdots,\mathbf{f}'_{BB,k}[I_k]\right]$. If the power constraint is satisfied, i.e., $\big\|\mathbf{F}_{RF,k}\mathbf{F}'_{BB,k}\big\|_F^2 \le P_k$, $\mathbf{F}_{BB,k} = \mathbf{F}'_{BB,k}$. Otherwise, the power constraint is imposed on $\mathbf{F}'_{BB,k}$ to produce a feasible baseband precoding solution as:
\begin{equation}\label{eqMLPbbF} % eq.16
	\mathbf{F}_{BB,k} = \frac{\mathbf{F}'_{BB,k}}{\sqrt{P_k}\big\|\mathbf{F}_{RF,k}\mathbf{F}'_{BB,k}\big\|_F}.
\end{equation}

For all the above MLPs ($p_{t,w}(\cdot),q_{t}(\cdot),MLP_{RF},MLP_{BB},t\in\mathcal{A},w\in\mathcal{E}$), batch normalization (BatchNorm) is utilized for the inputs to scale them into a similar range and the dropout strategy is adopted to avoid overfitting \cite{Ioffe,Srivastava}. The whole architecture of our proposed HGNN is shown in Fig.~\ref{GNN}, where the node update process is divided into BS nodes updating and UE nodes updating, both of which consist of aggregation and combination for BS nodes and UE nodes. Note that as each UE is served by only one BS, the attention mechanism in the aggregation for UEs through desired links is degraded into a direct link.

\subsection{Proposed HGNN based system operation}\label{S3.3}

The proposed HGNN based system adopts a centralized deployment, where a central processing unit (CPU) is attached to the sub-6GHz BS and responsible for the hybrid beamforming in the heterogeneous sub-6GHz and mmWave communication system. The CPU is equipped with the proposed HGNN and the parameters of the HGNN are trained offline and fine tuned online. 

In the training phase, each mmWave BS estimates the mmWave channel to each UE and transmits the mmWave CSI to the CPU in the sub-6GHz BS. Meanwhile, the sub-6GHz BS estimates the sub-6GHz channel to each UE. The HGNN in the CPU is trained with the both the mmWave and sub-6GHz CSI. The goal of the HGNN is to maximize the system's spectral efficiency, i.e., the problem (\ref{max}), an unsupervised learning strategy is adopted in the training phase with the loss function given by
\begin{equation}\label{loss} % eq.17
	Loss = -\sum_{k=1}^{K}\sum_{i=1}^{I_k}R_{i_k} .
\end{equation}
Adam algorithm \cite{Adam}, a gradient descent method, is used to train the parameters of the HGNN. The training procedure is summarized in Algorithm~\ref{alg}.

In the evaluation phase, the parameters of the HGNN and the selection of active mmWave antennas are fixed. The mmWave BSs estimate the partial mmWave CSI and the sub-6GHz BS estimates the sub-6GHz CSI. Both partial mmWave CSI and sub-6GHz CSI are transmitted to the CPU, which utilizes the well-trained HGNN to obtain the hybrid beamforming solution. The hybrid beamforming solution is transmitted to the mmWave BSs, which subsequently perform the hybrid beamforming and transmit the signals to the UEs.

\begin{algorithm}[b!]
\caption{Training of the proposed HGNN}
\label{alg} % Alg.1
\begin{algorithmic}[1]
	\Require Training dataset $\mathcal{D}$, number of iterations $MAX_{iter}$, number of GNN layers $L$, and initialized trainable parameters in $MLP_{BB}(\cdot)$, $MLP_{RF}(\cdot)$, $\mathbf{att}_{t,w}$, $p_{t,w}(\cdot)$, $q_{t}(\cdot)$, for $t\in\mathcal{A}$, $w\in\mathcal{E}$.
	\Ensure Well-trained parameters in $MLP_{BB}(\cdot)$, $MLP_{RF}(\cdot)$, $\mathbf{att}_{t,w}$, $p_{t,w}(\cdot)$, $q_{t}(\cdot)$, for $t\in\mathcal{A}$, $w\in\mathcal{E}$.
	\For{$n_{iter} =1: MAX_{iter}$}
	  \State Draw a random subset of $\mathcal{D}$.
	  \State Build heterogeneous graph with $v_{k}^{(0)} = P_k$, $\mathbf{v}_{i_k}^{(0)}=\big[\sigma^2_{i_k},\widetilde{\mathbf{h}}_{i_k}^{\rm T}\big]^{\rm T}$, $\mathbf{e}_{i_k,k}=\bar{\mathbf{h}}_{i_k,k}$.
	  \For{$l=1:L$} 
	    \State Update BS node features $\mathbf{v}_{k}^{(l)}$ with $\mathbf{v}_{k}^{(l-1)}$, $\mathbf{v}_{i_k}^{(l-1)}$ and $\mathbf{e}_{i_k,k}$ through aggregation and combination.
	    \State Update UE node features $\mathbf{v}_{i_k}^{(l)}$ with $\mathbf{v}_{k}^{(l-1)}$, $\mathbf{v}_{i_k}^{(l-1)}$ and $\mathbf{e}_{i_k,k}$ through aggregation and combination.
	  \EndFor
	  \State Compute analog precoder $\mathbf{F}_{RF,k}$ with $MLP_{RF}$.
	  \State Compute baseband precoder $\mathbf{F'}_{BB,k}$ with $MLP_{BB}$.
	  \For{$k=1:K$}
	    \If{$\|\mathbf{F}_{RF,k}\mathbf{F'}_{BB,k}\|_F^2 \le P_k$}
			  $\mathbf{F}_{BB,k} = \mathbf{F'}_{BB,k}$;
			\Else
		    \State Project $\mathbf{F'}_{BB,k}$ onto feasible space $\mathbf{F}_{BB,k} = \frac{\mathbf{F}'_{BB,k}}{\sqrt{P_k}\big\|\mathbf{F}_{RF,k}\mathbf{F}'_{BB,k}\big\|_F}$.
	    \EndIf
	  \EndFor
	  \State Compute loss function $Loss = -\sum_{k=1}^{K}\sum_{i=1}^{I_k}R_{i_k}$.
	  \State Use Adam or gradient descent method to update trainable parameters.
	\EndFor
\end{algorithmic}
\end{algorithm}

\subsection{Complexity analysis and overhead cost}\label{S3.4}

In the aggregation procedure, we have the MLPs $p_{t,w}(\cdot)$ with $t\in\{{\rm b},{\rm u}\}$ and $w\in\{{\rm d},{\rm i}\}$, while in the combination procedure, we have the MLPs $q_t(\cdot)$ with $t\in\{{\rm b},{\rm u}\}$. Let $C_{p_{{\rm b},{\rm d}}}$, $C_{p_{{\rm u},{\rm d}}}$, $C_{p_{{\rm b},{\rm i}}}$, $C_{p_{{\rm u},{\rm i}}}$, $C_{q_{\rm b}}$ and $C_{q_{\rm u}}$ denote the numbers of floating point operations (FLOPs) of the MLP functions $p_{{\rm b},{\rm d}}(\cdot)$, $p_{{\rm u},{\rm d}}(\cdot)$, $p_{{\rm b},{\rm i}}(\cdot)$, $p_{{\rm u},{\rm i}}(\cdot)$, $q_{\rm b}(\cdot)$ and $q_{\rm u}(\cdot)$, respectively. The FLOPs of an MLP are determined by the number of its layers and the number of neurons in each layer. Furthermore, let $C_{att_{{\rm b},{\rm d}}}$, $C_{att_{{\rm u},{\rm d}}}$, $C_{att_{{\rm b},{\rm i}}}$ and $C_{att_{{\rm u},{\rm i}}}$ denote the numbers of FLOPs in the corresponding attention mechanisms. In addition, Let $C_{BB}$ and $C_{RF}$ be the numbers of FLOPs in $MLP_{BB}(\cdot)$ and $MLP_{RF}(\cdot)$, respectively. Then the total number of FLOPs for the HGNN is expressed as
\begin{align}\label{eqComp} % eq.18
	C_{\rm total} =& L\Big( K \big( I_k \big(C_{p_{{\rm b},{\rm d}}} + C_{att_{{\rm b},{\rm d}}}\big) \nonumber \\
	&+ \big(I_{sum} - I_k\big) \big(C_{p_{{\rm b},{\rm i}}} + C_{att_{{\rm b},{\rm i}}}\big) + C_{q_{\rm b}}\big) \nonumber \\
	&+ I_{sum}\big(\big(C_{p_{{\rm u},{\rm d}}}\! +\! C_{att_{{\rm u},{\rm d}}}\big) \nonumber \\
	&+ (K\! -\! 1) \big(C_{p_{{\rm u},{\rm i}}}\! +\! C_{att_{{\rm u},{\rm i}}}\big) + C_{q_{\rm u}}\big) \Big) \nonumber \\
	&+ K C_{RF} + I_{sum} C_{BB}.
\end{align}
In practice, the number of layers $L$ for the HGNN is set to $2$ to $4$, and the total complexity is mainly dominated by the number of FLOPs in the aggregation part of the HGNN.

The overhead pilots consist of the pilots for sub-6GHz channel estimation and partial mmWave channel estimation. As the sub-6GHz CSI is needed for communication at sub-6GHz band, the extra pilot cost only includes the pilots for partial mmWave CSI estimation, which can be calculated as $K\cdot I_{sum} \cdot Pilots(\bar{N}_m)$, where $Pilots (\bar{N}_m)$ represents the number of pilots for $\bar{N}_m$ mmWave antennas.

The information required to be transmitted among different BSs consists of the partial mmWave CSI, the cost of which can be calculated as $K\cdot I_{sum} \cdot \bar{N}_m$, and the hybrid beamforming matrices, the cost of which can be calculated as $\sum_{k=1}^K (N_m \cdot I_{k} + I_{k} \cdot I_{k})$ for fully-connected beamforming and $\sum_{k=1}^K (N_m + I_{k} \cdot I_{k})$ for partially-connected beamforming. It is worth noting that information transmission between BSs is achieved through (wired) backhaul communication, resulting in relatively low communication costs. However, the actual implementation of such backhaul communication is challenging due to hardware costs and synchronization issues, especially in OFDM systems, where the CPU needs to send many precoding matrices to the BSs in a frequent way. Reducing the information exchange between BSs for the proposed HGNN is a challenging objective for the future study.

\section{Simulation results and discussions}\label{sim}

In this section, we evaluate the performance of our proposed HGNN.

\subsection{Simulation setup}\label{S4.1}

\begin{figure}[!t]
\vspace*{-2mm}
\begin{center}
\includegraphics[width=0.95\columnwidth]{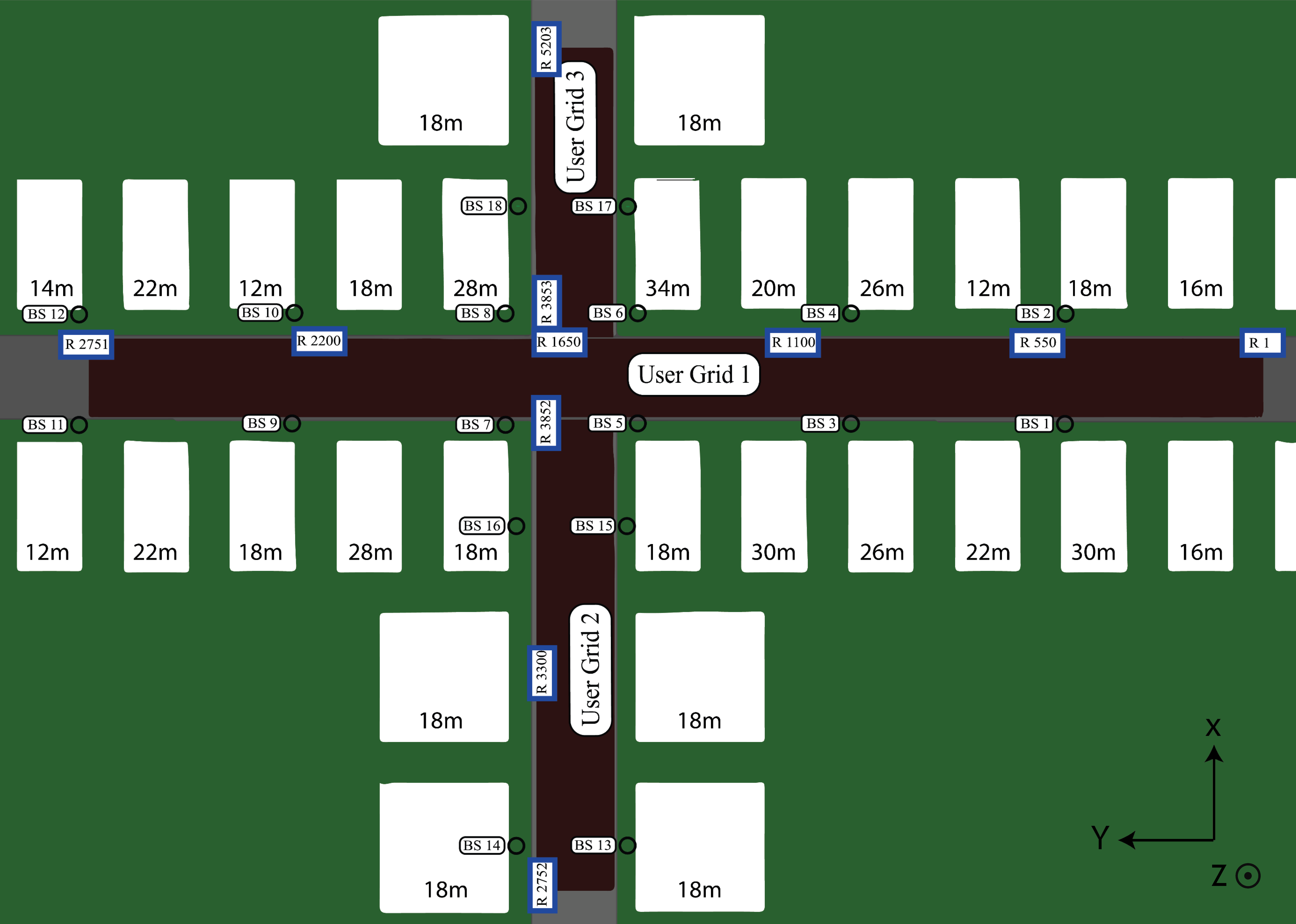}
\end{center}
\vspace{-2mm}
\caption{The `O1' scenario in \cite{Alkhateeb4}.}
\label{O1} % Fig.6
\vspace{-2mm}
\end{figure}

As shown in~Fig.~\ref{O1}, we use the `O1' scenario in the DeepMIMO dataset \cite{Alkhateeb4} to generate the sub-6GHz and mmWave channels. The DeepMIMO dataset is built by precise ray-tracing data from Remcom Wireless InSite and relays on the environment geometry/materials, which implies the reliability for machine learning. The sub-6GHz channel and the mmWave channel generated by DeepMIMO share the common environment, which determines the relevance of the sub-6GHz CSI and the mmWave CSI. 

\begin{table}[!t]
\vspace{-2mm}
\renewcommand{\arraystretch}{1.3}
\caption{Parameters to generate CSI in DeepMIMO dataset}
\vspace{-2mm}
\label{SimSetup} % Tab1
\begin{center}
\begin{tabular}{ccc}
	\hline
	Parameters        & mmWave          & sub-6GHz \\ \hline
	carrier frequency & 28GHz           & 3.5GHz \\
	activate BSs      & BS5 BS6 BS7 BS8 & BS5 \\
	BS antennas       & 32              & 8 \\
	Bandwidth         &100 MHz          & 10 MHz\\
	antenna spacing   & 0.5 wavelength  & 0.5 wavelength \\
	number of rays    &  25             & 25 \\ \hline
\end{tabular}
\end{center}
\vspace*{-2mm}
\end{table}

In the `O1' scenario, we divide the district from row $1400$ to row $2000$ in user grid 1 into $4$ cells. Each cell is served by one mmWave BS, and all the cells are served by one sub-6GHz BS. The number of active antennas for estimation of partial mmWave CSI is set to $\bar{N}_m=4$. $K=2,3,4$ BSs are set active and each BS serves $I_k=2,4$ UEs. The ULA antennas of both sub-6GHz and mmWave BSs are deployed in the y-axis direction. Other system parameters used in data generation are listed in Table~\ref{SimSetup}. We construct a dataset with $11,000$ independent samples, of which $10,000$ samples are used as training dataset and $1,000$ samples are used as test dataset. In each sample, the locations of UEs are randomly and uniformly generated in each cell. The sub-6GHz channels and mmWave channels are generated according to the locations of UEs by the DeepMIMO dataset.

\subsection{Parameters of the proposed HGNN}\label{S4.2}

For both fully-connected hybrid beamforming and partially-connected beamforming, the same network parameters are utilized except for the normalization layer for analog precoders $\mathbf{F}_{RF,k}$. The inputs of both heterogeneous graph networks are a structural heterogeneous graph data consisting of a graph $G$, a node type mapping function $\tau$ and a link type mapping function $\phi$, as defined in Subsection~\ref{hg}. In the following simulations, we use HGNN-FULLY to refer to the HGNN used for fully-connected hybrid precoding, and HGNN-PARTIALLY to refer to the HGNN used for partially-connected hybrid precoding. For the both HGNNs, we set $L=4$ and the detailed parameters are listed in Table~\ref{paraHGNN}

\begin{table}[t]
\vspace*{-2mm}
\renewcommand{\arraystretch}{1.3}
\caption{Parameters of the proposed HGNN and training algorithm}
\vspace{-2mm}
\label{paraHGNN} % Tab2
\vspace*{-2mm}
\begin{center}
\begin{tabular}{ccc}
	\hline
	Layers $L$ & 4 \\
	Batch size & 10 \\
	Learning rate & 0.001, $\times0.9$ every 5 epoch \\
	Sample number of training set & 10000 \\
	Sample number of test set & 1000 \\
	Epoch & 50 \\
	Dropout & 0.3 \\
	Dimension of node features & 600 \\
	$p_{t,w}(\cdot),t\in\mathcal{A},w\in\mathcal{E}$ & (608,800,600) \\
	$q_{t}(\cdot),t\in\mathcal{A}$ & (1200,800,600) \\
	$MLP_{RF}$ & (600,600,64) \\
	$MLP_{BB}$ & (1208,50,4) \\ \hline
\end{tabular}
\end{center}
\vspace*{-2mm}
\end{table}

\subsection{Evaluation of the proposed HGNN}\label{S4.3}

\subsubsection{Convergence}

\begin{figure}[!t]
\vspace*{-2mm}
\begin{center}
\includegraphics[width=0.8\columnwidth]{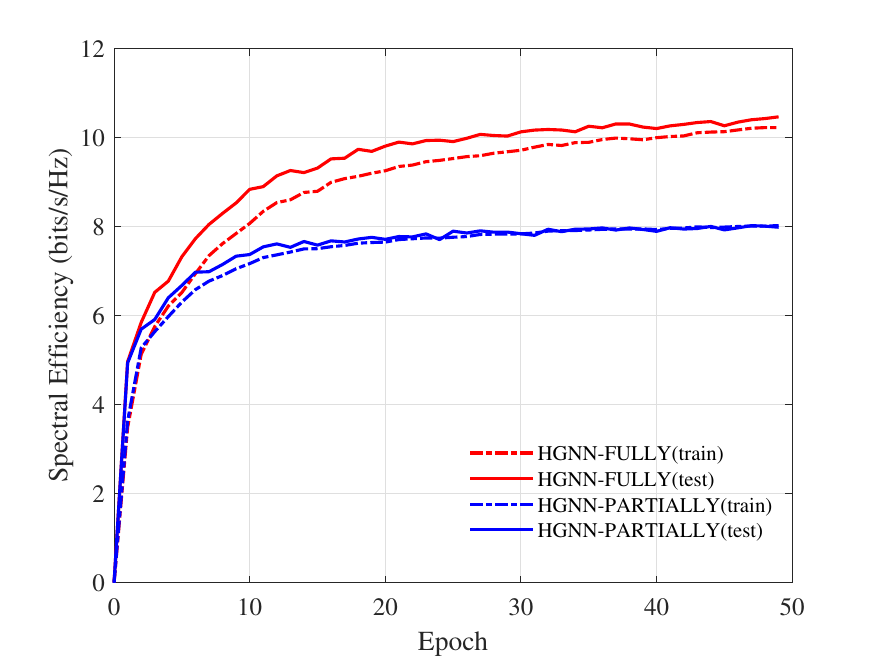}
\end{center}
\vspace{-2mm}
\caption{Training and test learning curves for HGNN-FULLY and HGNN-PARTIALLY, in terms of spectral efficiency.}
\label{epoch} % Fig.7
\vspace*{-2mm}
\end{figure}

We first evaluate the convergence of our HGNNs, and Fig.~\ref{epoch} depicts the spectral efficiency learning curves of HGNN-FULLY and HGNN-PARTIALLY. It can be seen that the training of HGNN-FULLY converges at about $40$-th epoch while HGNN-PARTIALLY converges at about $20$-th epoch. As expected, HGNN-PARTIALLY converges faster than HGNN-FULLY due to fewer outputs of the former. It is interesting to see that the test spectral efficiency on the test set is slightly higher than the training spectral efficiency for both HGNN-FULLY and HGNN-PARTIALLY, which is owing to \emph{Dropout}\footnote{Dropout works by randomly dropping out (i.e., setting to zero) some of the neurons in a neural network during training. This is a common technique to assist a deep neural network avoiding overfitting and help improving the generalization, i.e., improving the performance on the data unseen in training. Although the random dropout of neurons may miss some information over connection of neurons, which would cause the training performance `degradation', this kind of regularization can often avoid overfitting into the features only related to the training data, e.g., noise, and therefore enhances the model generalization capability. This reflects in Fig.~\ref{epoch} where the test spectral efficiency is slightly higher than the training spectral efficiency.}. In addition, the fully-connected structure achieves higher spectral efficiency than the partially-connected structure as expected.

\begin{figure}[t!]
\vspace*{-2mm}
\begin{center}
\includegraphics[width=0.8\columnwidth]{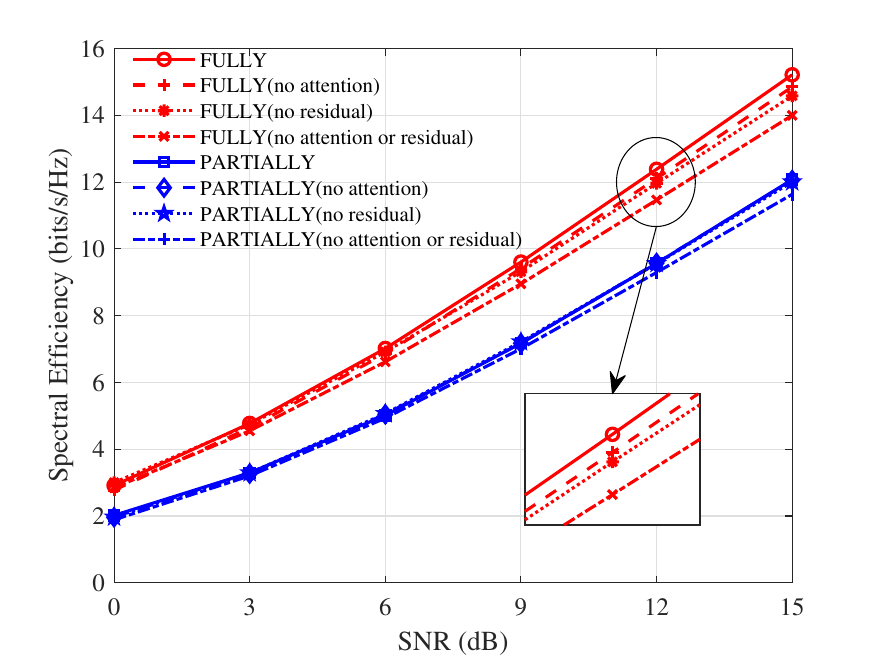}
\end{center}
\vspace{-2mm}
\caption{Impact of attention mechanism and residual structure on the achievable performance of HGNN.}
\label{attres} % Fig.8
\vspace*{-2mm}
\end{figure}

\begin{figure}[t!]
\vspace*{-2mm}
\begin{center}
\includegraphics[width=0.8\columnwidth]{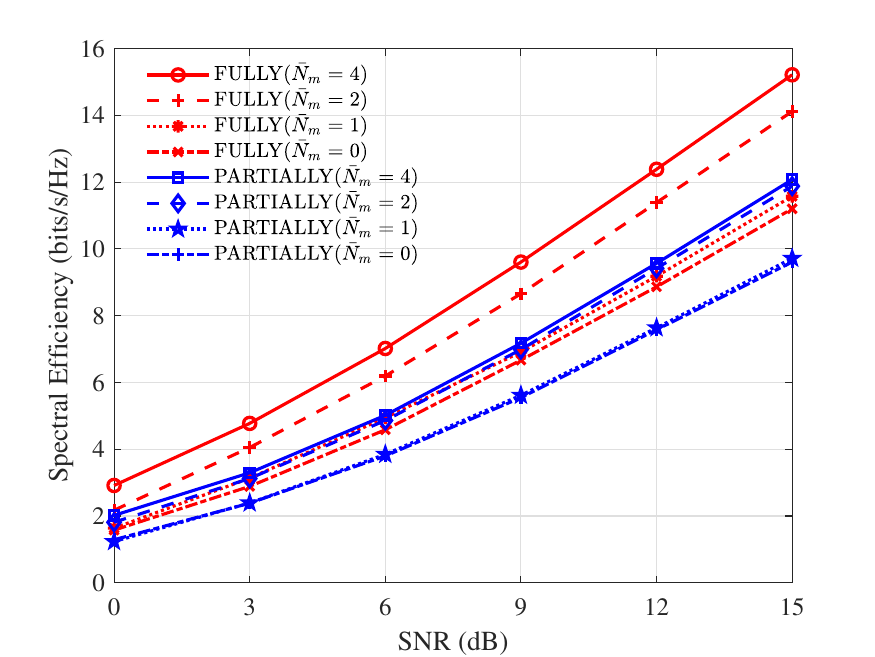}
\end{center}
\vspace{-2mm}
\caption{The spectral efficiency comparison of HGNN with various numbers of active antennas $\bar{N}_m$, through which partially mmWave CSI is estimated.}
\label{beam} % Fig.9
\vspace*{-2mm}
\end{figure}

\subsubsection{Effectiveness of attention mechanism and residual structure}

Fig.~\ref{attres} investigates the impact of attention mechanism and residual structure on the achievable performance of the HGNN. The aggregation procedure of the HGNN without attention is obtained by replacing (\ref{aggWat}) with $\mathbf{a}_{i,t,w}^{(l)} = \sum_{j\in\mathcal{N}_{t,w}(\mathbf{v}_i)} p_{t,w}(\mathbf{v}_j^{(l-1)},\mathbf{e}_{i,j})$, while the combination procedure of the HGNN without residual structure is obtained by replacing (\ref{combWres}) with $\mathbf{v}_i^{(l)}=q_{\tau(\mathbf{v}_i)}\left(\mathbf{v}_i^{(l-1)},\{\mathbf{a}_{i,t,w}^{(l)},t\in\mathcal{A},w\in\mathcal{E}\}\right)$. It can be seen that the performance gain of the attention mechanism and the residual structure is more significant in the fully-connected structure than in the partially-connected structure. This is because the optimization problem  (\ref{max}) for the fully-connected structure is more complicated, and there are more degrees of freedom in design to be exploited by the attention mechanism and the residual structure in the fully-connected structure. The results of Fig.~\ref{attres} hence verify the effectiveness of attention mechanism and residual structure.

\subsubsection{The effect of amount of partial mmWave CSI}

The amount of partial mmWave SCI is measured by $\bar{N}_m$, the number of active antennas in the partial mmWave CSI estimation procedure. More active antennas means more information about the mmWave channel. Fig~\ref{beam} shows the performance of the HGNN with different numbers of active antennas $\bar{N}_m$. It can be seen that the HGNN works even with only sub-6GHz CSI and no partial mmWave CSI. As expected, with more partial mmWave CSI, the HGNN achieves higher spectral efficiency, at the expense of higher training overhead. It can also be observed that the performance gain from increasing the partial mmWave CSI is more significant for the fully-connected structure than for the partially-connected structure.

\begin{figure}[t!]
\vspace*{-2mm}
\begin{center}
\includegraphics[width=0.8\columnwidth]{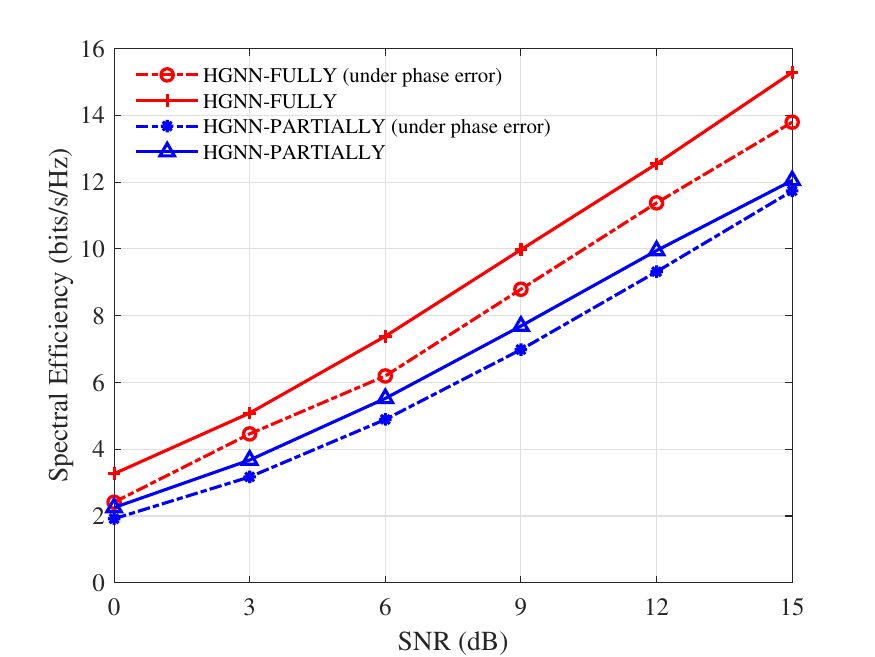}
\end{center}
\vspace{-2mm}
\caption{The spectral efficiency of HGNN with/without the random phase error of CSI.}
\label{phaseerror} % Fig.10
\vspace*{-2mm}
\end{figure}

\begin{figure}[t!]
	\vspace*{-2mm}
	\begin{center}
		\includegraphics[width=0.8\columnwidth]{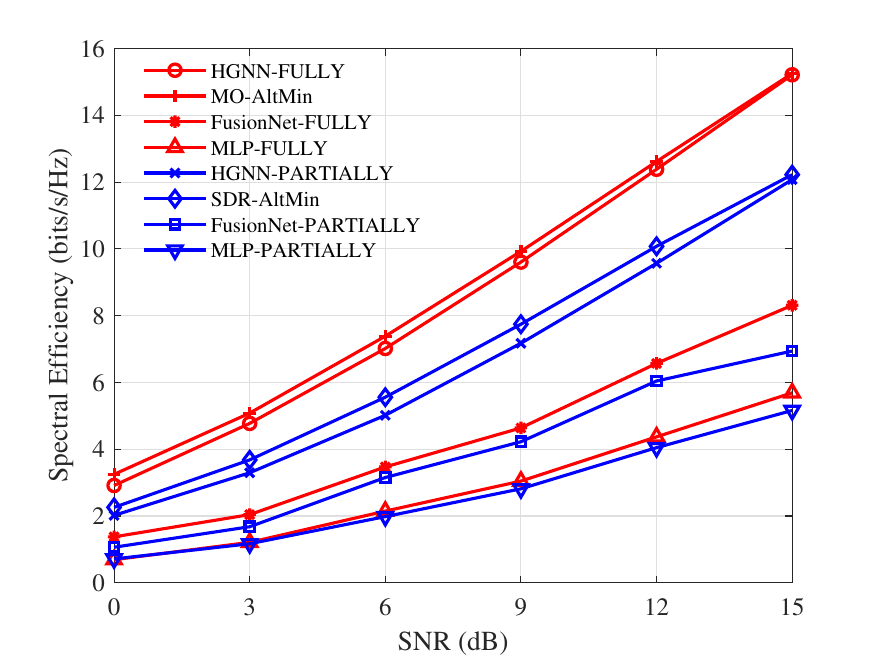}
	\end{center}
	\vspace{-2mm}
	\caption{The spectral efficiency of HGNN, MLP, FusionNet and alternating methods as the functions of SNR.}
	\label{methods} % Fig.11
	\vspace*{-2mm}
\end{figure}

\subsubsection{The robustness of the proposed HGNN under phase error}

To evaluate the robustness of our proposed method, we further test the performance of our HGNN under the CSI with random phase error. The random phase error is generated by $\mathbf{e}_{i,j} = \mathbf{e}_{i,j} \cdot e^{\textsf{j}\theta}$, where $\theta$ is a random variable, following the normal distribution $\mathcal{N}(0,(5^\circ)^2)$. Fig.~\ref{phaseerror} shows the spectral efficiency of our HGNN with/without the random phase error of CSI. It can be seen that the performance of the HGNN degrade slightly due to the random phase error, which implies that our HGNN is robust to the random phase error of CSI.

\begin{table*}[t!]
	\vspace*{-2mm}
	\renewcommand{\arraystretch}{1.3}
	\caption{Comparison of test spectral efficiency (bits/s/Hz) for different methods with different numbers of UEs and mmWave BSs.}
	\label{generalization} % Tab 3
	\vspace{-2mm}
	\begin{center}
		\begin{tabular}{ccccccc}
			\hline
			& HGNN-FULLY & HGNN-PARTIALLY & MLP-FULLY & MLP-PARTIALLY & MO-AltMin & SDR-AltMin \\ \hline
			2BSs 2UEs  & 10.2 & 8.2 & 4.1 & 3.8 & 11.2 & 8.8 \\
			2BSs 8UEs  & 16.5 & 10.6 & 4.3 & 4.3 & 18.7 & 9.0 \\
			3BSs 6UEs  & 16.0 & 12.0 & 5.1 & 4.1 & 16.8 & 12.9 \\
			3BSs 12UEs & 25.1 & 15.7 & 4.2 & 4.3 & 27.6 & 12.8 \\
			4BSs 8UEs  & 21.0 & 15.7 & 5.3 & 5.2 & 21.7 & 16.6 \\
			4BSs 16UEs & 32.0 & 20.2 & 4.0 & 3.7 & 35.0 & 15.8 \\ \hline
		\end{tabular}
	\end{center}
	\vspace*{-2mm}
\end{table*}

\begin{table*}[t!]
	\vspace*{-2mm}
	\renewcommand{\arraystretch}{1.3}
	\caption{Comparison of run time (s) for different schemes with different numbers of UEs and mmWave BSs.}
	\vspace{-2mm}
	\label{time} % Tab 4
	\begin{center}
		\begin{tabular}{ccccccc}
			\hline
			& HGNN-FULLY & HGNN-PARTIALLY & MLP-FULLY & MLP-PARTIALLY & MO-AltMin & SDR-AltMin \\ \hline
			2BSs 4UEs&58.36 &2.08 &0.22 &0.26 &96.87 &420.22 \\
			2BSs 8UEs&62.06 &2.35 &0.23 &0.25 &253.29 &939.62 \\
			3BSs 6UEs&65.17 &1.95 &0.13 &0.26 &136.76 &562.46 \\
			3BSs 12UEs & 65.08 & 2.18 & 0.24 & 0.26 & 356.03 & 1452.97 \\
			4BSs 8UEs  & 58.38 & 2.52 & 0.24 & 0.25 & 178.71 & 775.45 \\
			4BSs 16UEs & 61.65 & 2.36 & 0.24 & 0.34 & 629.99 & 1900.43 \\ \hline
		\end{tabular}
	\end{center}
	\vspace*{-2mm}
\end{table*}

\begin{table*}[t!]
	\vspace{-2mm}
	\renewcommand{\arraystretch}{1.3}
	\caption{Overhead comparison for different schemes.}
	\vspace{-2mm}
	\label{overhead} % Tab 4
	\begin{center}
		\begin{tabular}{c c c c}
			\hline
			& HGNN & MLP/FusionNet & MO/SDR-AltMin \\ \hline
			mmWave pilot overhead&$K\cdot I_{sum}\cdot Pilots(\bar{N}_m)$ &$ K\cdot I_{sum}\cdot Pilots(\bar{N}_m)$ &$K\cdot I_{sum}\cdot Pilots(N)$ \\ 
			Backhaul overhead (fully) &$K\cdot I_{sum}\cdot \bar{N}_m + \sum_{k=1}^{K}(N_m\cdot I_k + I_k^2)$ &0 &$K\cdot I_{sum}\cdot N_m + \sum_{k=1}^{K}(N_m\cdot I_k + I_k^2)$ \\ 
			Backhaul overhead (partially) &$K\cdot I_{sum}\cdot \bar{N}_m + \sum_{k=1}^{K}(N_m + I_k^2)$ &0 &$K\cdot I_{sum}\cdot N_m + \sum_{k=1}^{K}(N_m + I_k^2)$\\ \hline
		\end{tabular}
	\end{center}
	\vspace{-2mm}
\end{table*}

\subsection{Performance comparison}\label{S4.4}

\subsubsection{Comparison benchmarks}

We compare our method with the MLP method, FusionNet method \cite{Gao} and the alternating methods, i.e., MO-AltMin for fully-connected structure and SDR-AltMin for partially-connected structure \cite{XYu}.

In the MLP method, a vanilla MLP is utilized to generate both analog beamforming matrices $\mathbf{F}_{RF,k}$ and digital matrices $\mathbf{F}_{BB,k}$, which are subsequently normalized to satisfy the constant modulus constraint and power constraint. Similarly, MLP-FULLY refers to the MLP for fully-connected hybrid precoding while MLP-PARTIALLY refers to the MLP for partially-connected hybrid precoding. The number of hidden layers for the both MLPs is $3$, and the numbers of neurons in the three hidden layers are $200,300,500$, respectively. Since MLPs cannot directly deal with heterogeneous graph data, all the node features, edge features, node types, and edge types are concatenated into a vector as the input of MLPs. For fairness, both MLPs are trained in an unsupervised manner with the same loss function (\ref{loss}) as our HGNNs.

In the FusionNet method, we adopt the same network structures as FusionNet in \cite{Gao}, where the sub-6GHz information and mmWave information are fused in an attention layer of the network. The original FusionNet is designed for beam selection in the heterogeneous network where sub-6GHz and mmWave coexist, and we modify its outputs as hybrid precoding matrix. Similarly, the FusionNet-FULLY refers to the FusionNet for fully-connected hybrid precoding while FusionNet-PARTIALLY refers to the FusionNet for partially-connected hybrid precoding. The number of hidden layers for the both FusionNets is $4$, and the dimensionality of hidden features for both sub-6GHz and mmWave before fusion is $300$. FusionNet-FULLY and FusionNet-PARTIALLY are trained in an unsupervised manner with the same loss function (\ref{loss}).

In \cite{XYu}, the alternating methods were introduced to approximate the optimal fully digital precoding matrix in the metric of Frobenius norm by alternately iterating between the baseband precoding matrix and the analog precoding matrix based on the traditional optimization theory. Specifically, the MO-AltMin is an optimization algorithm based on the manifold optimization theory for fully-connected hybrid beamforming, while the SDR-AltMin is based on the semi-definite relaxation (SDR) algorithm for partially-connected hybrid beamforming. The optimal fully digital precoding matrices for the MO-AltMin and SDR-AltMin are generated by the WMMSE method \cite{Shi} in the simulation. Note that unlike the MLP method and our HGNN, the entire mmWave CSI is required for the both MO-AltMin and SDR-AltMin methods.

\subsubsection{Comparison for different SNRs}

We first compare the performance of the HGNN, FusionNet, MLP and traditional alternating methods at different SNR levels in Fig.~\ref{methods}. It can be seen that the performance of HGNN-FULLY and HGNN-PARTIALLY are very close to the performance of MO-AltMin and SDR-AltMin, respectively, over the whole SNR range tested. This is very significant considering that only sub-6GHz and partially mmWave CSI are fed to the HGNNs, while the whole mmWave CSI is fed to the alternating methods. This result also implies that the HGNN can learn the relationships between the sub-6GHz channel and the mmWave channel well and is capable of solving the challenging precoding optimization problem (\ref{max}) effectively. Also observe that the performances of the HGNNs are significantly better than the FusionNets and MLPs, due to the capability of HGNN to deal with heterogeneous graph data and interference between UEs.

\subsubsection{Comparison for scalability of different BSs and UEs}

We further test the scalability of the proposed HGNN, MLP and alternating methods by adjusting the number of mmWave BSs and the number of UEs per mmWave BS served, and the results are presented in Table~\ref{generalization}. It can be seen that the performance of HGNN-FULLY remains close to the performance of MO-AltMin method for all the configurations. It can also be seen that the performance of HGNN-PARTIALLY are even better than the performance of SDR-AltMin method in some configurations, especially when there are more UEs per mmWave BS served. Furthermore, the spectral efficiency gap between HGNN and MLP methods increases as the number of BSs and UEs increase. This is because the MLP methods are unable to effectively deal with the interference between UEs and have low scalability for the network size.

\subsubsection{Comparison of run time}

We run MO-AltMin and SDR-AltMin on \emph{$12$th Gen Intel (R) Core(TM) i7-12700KF} with 20 processors while the MLPs and the proposed HGNNs on \emph{GeForce RTX 3080 Ti} to demonstrate the computational advantages of the proposed HGNN over the traditional alternating methods. Note that both MO-ALtMin and SDR-AltMin are unable to exploit the parallel computation of GPU due to their sequential computational flows. The total run times of various methods on the test dataset are compared in Table~\ref{time}. It is obvious that the proposed HGNNs consume much less run time than the alternating methods. In addition, the run times of the HGNNs are almost constant with different numbers of BSs and UEs, while the run times of MO-AltMin and SDR-AltMin increase as the numbers of BSs and UEs increase. The MLP methods impose the lowest run time but their performance are the worst.

\subsubsection{Comparison of overhead}

The total communication overhead consists of the additional mmWave pilot overhead for channel estimation and the backhaul overhead for information exchange between BSs. The comparison of overhead for different schemes is shown in Table~\ref{overhead}. It can be seen that HGNN has the same pilot overhead as MLP and FusionNet methods, while the MO/SDR-AltMin methods require the whole CSI, resulting in higher pilot overhead. The backhaul overhead of HGNN is lower than that of MO/SDR-AltMin methods, as HGNN only requires the partial CSI to be exchanged between BSs, while MLP and FusionNet methods do not require the information exchange between BSs, and hence have no backhaul overhead.

\vspace{-3mm}
\section{Conclusions}\label{clu}

In this paper, we have focused on the multi-user multi-cell hybrid beamforming problem at mmWave band under the co-existence of sub-6GHz and mmWave communications. In order to reduce the overhead cost, the sub-6GHz CSI and only partial mmWave CSI have been utilized. We have modeled the system as a heterogeneous graph and have proposed a HGNN with the attention mechanism and residual structure to solve this challenging problem. The numerical results have demonstrated that our proposed HGNN outperforms other machine learning methods in various scenarios. Moreover, the utilization of attention mechanism and residual structure has been shown to enhance the performance of the proposed HGNN. For the future work, it would be interesting to investigate the reduction of the communication overhead between BSs in more complex scenarios including the non-orthogonal multiple access and RIS aided systems.
\vspace{-3mm}

\end{document}